\documentclass[12pt]{article}

\title{Chiral Symmetry Breaking in QCD}
\author{  Yu.A.Simonov\\
 State Research Center\\Institute of Theoretical and Experimental
Physics, \\ Moscow, 117218 Russia}
 \date{}
\newcommand{\beq}{\begin{eqnarray}}
 \newcommand{\eeq}{\end{eqnarray}}
\newcommand{\be}{\begin{equation}}
 \newcommand{\ee}{\end{equation}}

\def\fun#1#2{\lower3.6pt\vbox{\baselineskip0pt\lineskip.9pt
\ialign{$\mathsurround=0pt#1\hfil ##\hfil$\crcr#2\crcr\sim\crcr}}}

\newcommand{{\SD}}{\rm SD}

\newcommand{\vex}{\mbox{\boldmath${\rm x}$}}

\newcommand{\ver}{\mbox{\boldmath${\rm r}$}}

\newcommand{\lan}{\langle}
\newcommand{\ran}{\rangle}
\begin{document}
\maketitle

\begin{abstract}
Chiral Symmetry Breaking (CSB) is derived in QCD starting from the
QCD Lagrangian and using Field Correlators Method (FCM). The
kernel in the resulting equations responsible for CSB is directly
connected to confinement, and therefore both phenomena occur and
vanish together as supported by lattice data.

Chiral Lagrangian and quark-meson Lagrangian are derived with
explicit coefficients and compared to standard expressions.
Spectrum of Nambu-Goldstone mesons and their radial excitations is
calculated in good agreement with experiment.

 \end{abstract}

 \section{Introduction}
Chiral Symmetry Breaking (CSB) is known to govern the low-energy
properties of hadrons \cite{1}. The Effective Chiral  Lagrangians
(ECL) have been proposed \cite{2,3} before the advent  of QCD, and
the phenomenon  of CSB and Nambu-Goldstone theorem was established
more than 40 years ago \cite{4}.

When the QCD era began, it was natural to ask whether CSB can be
derived from the QCD Lagrangian. The direct way being difficult,
some indirect arguments have been found in favor of CSB.  In 1971
Dolgov   and Zakharov \cite{5} and  in 1980 Coleman and Witten
\cite{6} noticed that the analytic  properties of the anomalous
triple correlator of currents cannot be ensured unless there is a
pseudoscalar pole (massless in the chiral limit). This signifies
CSB at least in the large $N_c$ limit, but tells nothing about the
mechanism of CSB and field  configurations responsible for it. For
the latter instantons  have been proposed and studied in numerous
papers. It was shown that instantons indeed can explain CSB both
qualitatively and quantitatively, if Effective Quark Lagrangian
(EQL) is taken in the  'tHooft form and instanton density
corresponds to the gluon condensate value \cite{8}. Another
evidence which seemingly favored instanton model, is the quark
mode distribution, which should be nonzero for small eigenvalues
\cite{9}  and was measured  on the  lattice (see \cite{10} and
refs therein).  On more theoretical ground the instantons have
been shown to create the stochastic matrix ensemble for
eigenvalues \cite{11}, and  the so-called Wigner's semicircle for
the spectrum which automatically ensures CSB.

The measurements however revealed a more sophisticated picture
with correlations not specific for  instantons (see \cite{12} and
refs therein) so the dynamics of CSB should be rather complicated.
In addition the ansatz of the 'tHooft determinant for the EQL in
the instanton model was seriously questioned \cite{13} and more
complicated EQL was computed directly in the model \cite{14,15}
with yet unknown properties.

The most serious drawback of instanton model is that it does not
ensure confinement, hence it may be responsible for only one part
of total picture. Therefore in \cite{15} another model was
suggested, containing instantons together with background vacuum
fields ensuring confinement. The model was shown to give a
reasonable qualitative picture, but  instantons are essential
element producing CSB with density comparable to the confining
background. However in \cite{16} it was found that the Casimir
scaling for charges in different SU(3) representations strongly
limits the density of instantons, thus making them not the chief
source of CSB. Therefore background fields themselves should
ensure CSB, and what was found recently in this field is
 reported below. In 1997 it was proved \cite{17} that CSB occurs
 in the example of heavy-light system due to the same field
 configurations (field correlators) which create the confining
 string. Recently \cite{18,20} the dynamics  of CSB was studied in
 more general context, and ECL \cite{18}, Nambu-Goldstone spectrum
 \cite{19} and  CSB order parameters \cite{20} have been obtained from the
 QCD Lagrangian without extra parameters. It is a purpose of the
 present talk to give a brief outline of the derivation and
 results of these investigations.  The plan of the talk is as
 follows. In section 2 the derivation of the Effective Quark
 Lagrangian (EQL), Effective Quark-Meson  Lagrangian (QML) and ECL
 are given, in section 3 properties of those are discussed and the
  Gell-Mann-Oakes-Renner (GOR) relations are derived, as well as
  the spectrum of Nambu-Goldstone mesons and their radial
  excitations. In section 5 the role of QML in the description of
  the decays and channel coupling is described, the concluding
  section summarizes the results.

  \section{Derivation of Effective Lagrangian}

  One starts with the QCD partition function in the Euclidean
  space-time.

  \be
Z=\int DAD\psi D\psi^+ e^{-S_0(A)+\int~^f\psi^+(i\hat
\partial+im+g\hat A)~^f\psi d^4x}
\label{1}\ee where $S_0(A)=\frac{1}{4}\int(F^a_{\mu\nu}(x))^2
d^4x$, $m$ is the current quark mass (mass matrix $\hat m$ in
SU(3)), and the quark operator $^f\psi_{a\alpha}(x)$ has flavor
index $a(f=1,... n_f)$, color index $a(a=1, ... N_c)$ and Lorenz
bispinor index $\alpha(\alpha=1,2,3,4)$, and use the contour gauge
to express $A_\mu(x)$ in terms of $F_{\mu\nu}$.
 One has for the contour $z_\mu(s,x)$ starting at point $x$ and ending at
$Y=z(0,x)$
\be
A_\mu(x)=\int^1_0 ds\frac{\partial z_\nu(s,x)}{\partial
s}\frac{\partial z_\rho(s,x)}{\partial x_\mu} F_{\nu\rho}
(z(s))\equiv \int^x_Y d \Gamma_{\mu\nu\rho} (z) F_{\nu\rho} (z)
.\label{2} \ee Integrating out the gluonic fields $A_{\mu} (x)$,
one obtains
\be
Z=\int D\psi D\psi^+ e^{\int~^f\psi^+(i\hat \partial+im)^f\psi
d^4x} e^{L^{(2)}_{EQL}+L^{(3)}_{EQL}+...}\label{3} \ee where the
EQL proportional to $\lan\lan A^n\ran\ran$ is denoted by
$L_{EQL}^{(n)}$, \be L^{(2)}_{EQL}=\frac{g^2}{2}\int
d^4xd^4y~^f\psi^+_{a\alpha}(x)
~^f\psi_{b\beta}(x)~^g\psi^+_{c\gamma}(y)~^g\psi_{d\varepsilon}(y)
\lan  A^{(\mu)}_{ab}(x) A^{(\nu)}_{cd}(y)\ran
\gamma^{(\mu)}_{\alpha\beta} \gamma^{(\nu)}_{\gamma\varepsilon}
\label{4} \ee Average of gluonic fields can be computed using
(\ref{2}) as
\be
g^2\lan A^{(\mu)}_{ab}(x) A^{(\nu)}_{cd}(y)\ran=
\frac{\delta_{bc}\delta_{ad}}{N_c} \int^x_0
du_i\alpha_{\mu}(u)\int^y_0 dv_k\alpha_\nu(v)
D(u-v)(\delta_{\mu\nu}\delta_{ik}-\delta_{i\nu}\delta_{k\mu}),
\label{5} \ee  where $D(x)$ is the correlator $\lan F(x)
F(0)\ran$. As it was argued in \cite{17} the dominant contribution
at large distances from
 the static antiquark is given by the color-electric fields, therefore we shall write down
 explicitly $L_{EQL}^{(2)} (el)$ for this case, i.e. taking $\mu=\nu=4$. As a result one has
\be
L_{EQL}^{(2)}(el)=\frac{1}{2N_c}\int d^4x\int
d^4y~^f\psi^+_{a\alpha}(x)~^f\psi_{b\beta}(x)
~^g\psi^+_{b\gamma}(y)~^g\psi_{a\varepsilon}(y)
\gamma^{(4)}_{\alpha\beta}\gamma^{(4)}_{\gamma\varepsilon} J(x,y)
\label{6} \ee where $J(x,y)$ is
\be
J(x,y) =\int^x_0 du_i\int^y_0 dv_i D(u-v),~~i=1,2,3. \label{7} \ee
One can form bilinears  $\Psi^{fg}_{\alpha\varepsilon}\equiv
~^f\psi^+_{a\alpha}~^g\psi_{a\varepsilon}$ and project using Fierz
procedure given isospin and Lorentz structures,
$\Psi^{fg}_{\alpha\varepsilon}\to \Psi^{(n,k)}(x,y).$ With the
help of the standard bosonization trick
\be
e^{-\Psi\tilde J\Psi}= \int(\det \tilde J)^{1/2} D\chi\exp [-\chi
\tilde J\chi+ i\Psi\tilde J \chi + i\chi \tilde J \Psi] \label{8}
\ee
\be
Z=\int D\psi D\psi^+ D\chi \exp L_{QML} \label{9} \ee
 one obtains the effective Quark-Meson Lagrangian (QML)
$$ L^{(2)}_{QML} =\int d^4x\int
d^4y\left\{~^f\psi^+_{a\alpha}(x)[(i\hat\partial+im)_{\alpha\beta}\delta(x-y)
+iM^{(fg)}_{\alpha\beta} (x,y)]~^g\psi_{a\beta}(y)- \right.$$
\be
\left.-\chi^{(n,k)}(x,y)\tilde J(x,y) \chi^{(n,k)}(y,x)\right\}
\label{10} \ee and the effective  quark-mass operator is
\be
M^{(fg)}_{\alpha\beta}(x,y) =\sum_{n,k} \chi^{(n,k)}(x,y)\bar
O^{(k)}_{\alpha\beta}t^{(n)}_{fg}\tilde J(x,y). \label{11} \ee

The QML in Eq.(\ref{10}) $L^{(2)}_{QML}$ contains  functions $
\chi^{(n,k)}$  which are integrated out in (\ref{11}), and the
standard way is to find $\chi^{(n,k)}$ from the  stationary point
of $L^{(2)}_{QML}$. Limiting oneself to the scalar and
pseudoscalar fields and using the nonlinear parametrization one
can write for the operator $\hat M$ in (\ref{10})
\be
\hat M(x,y)=M_{S}(x,y) \hat U(x,y),\hat
U=exp(i\gamma_{5}\hat\phi), \hat \phi(x,y)=\phi^{f}(x,y) t^{f}.
\label{12}\ee After integrating out the quark fields one obtains
the ECL in the form
\be
L^{(2)}_{ECL}(M_S,\hat \phi)=-2n_f (\tilde J (x,y))^{-1}
M^2_S(x,y)+ N_c tr\log[(i\hat\partial+im)\hat 1+iM_S \hat
U].\label{13}\ee The stationary point equations $\frac{\delta
L^{(2)}_{ECL}}{\delta M_s}= \frac{\delta L^{(2)}_{ECL}}{\delta
\hat \phi}=0$ at $\hat \phi=\hat\phi_0$, $M_s=M_s^{(0)}$
immediately show that $\hat \phi_0=0$ and $M^{(0)}_s$ satisfies
nonlinear equation
\be
 i M^{(0)}_{S}(x,y)=\frac{N_c}{4} tr S \tilde J(x,y)=
 N_c (\gamma_4 S \gamma_4) \tilde J(x,y) ,~~S(x,y)=-[i\hat\partial+im +iM_S \hat U]^{-1}_{x,y}.
 \label{14}\ee
The solution of (\ref{14}) was studied in \cite{17} and it was
shown that in the limit of small $T_g$
 one obtains for $M_{S}(x,y)$ a localized
 expression
 \be
 M_{S}(x,y) \approx \sigma|\vex| \delta^{(4)}(x-y),|\vex|\gg T_{g}. \label{15}
 \ee
 The ECL (\ref{13}) with the operator  $M_s$ in (\ref{15}) signals
 both confinement and CSB which  create  the
 Nambu-Goldstone meson spectrum, discussed in the next section.

\section{Nambu-Goldstone spectrum  from ECL}

Expanding ECL (\ref{13}) in the powers of  field  $\hat \phi$ one
arrives at the expression
\be
W^{(2)}(\phi) =\frac{N_c}{2} \int \phi_a (k) \phi_a(-k) \bar N(k)
\frac{d^{4}k}{(2\pi)^4}\label{16}\ee with
\be
\bar N(k)= \frac12 tr\{(\Lambda_+ M_S)_0+ \int
d^{(4)}ze^{ikz}\Lambda_+(0,z) M_S(z) \Lambda_-(z,0)
M_S(0)\}\label{17}\ee and $\Lambda_{\pm}=(\hat \partial\pm m\pm
M_s)^{-1}$.

 The pion mass is proportional to $\bar N(0)$, which
can be written as \be \bar{N}(0)=\frac12 tr (\Lambda_+
M_s\Lambda_- (\hat \partial-m))=\frac{m}{2} tr \Lambda_+
=-\frac{m}{4 N_c} \lan \bar \psi \psi \ran.\label{18}\ee
 Taking into account that $\phi_a=\frac{2\pi_a}{f_\pi},~~ f_\pi =
 93$ MeV, one obtains GOR relations
 \be
 2m^2_\pi f^2_\pi =  (m_u+m_d) |\bar \psi \psi|, ~~  |\bar \psi \psi|= |\bar u u| + |\bar
 d d|,\label{19}\ee
 and similar relations for  $m^2_k, m^2_\eta$ \cite{19}.
 An interesting question now is what happens with the pion radial
 excitations, and this will bring us to the point of connection
 between quark-model poles and Nambu-Goldstone poles in the total
 PS Green's function $G_{ab}$ -- correlator of the currents
 $J^{(5)}_a(x) =\bar\psi(x) \gamma_5 t_a \psi(x)$. At large $N_c$
 one can expand in powers of $\hat
 \phi=\frac{2\varphi_at_a}{f_\pi}=O\left(\frac{1}{\sqrt{N_c}}\right)$
 and keep the terms up to $O(\hat \phi^2)$ which allows to express
 $G_{ab}$ in terms of the function $G^{(0)}_{ab} =tr [S(x,y)
 \gamma_5 t_bS(y,x)\gamma_5t_a]$ and $G_{ab}^{(M)}$ and
 $G_{ab}^{(MM)}$ which contain one or two operators $M_S$ as
 factors of $\gamma_5 t_a, \gamma_5 t_b$. Now each of the function
 $G^{(0)}, G^{(M)}, G^{(MM)}$ can be computed using spectral
 representation in the quark model without chiral degrees of
 freedom, e.g.
 \be G^{(0}{(k)}= \sum^\infty_{n=0} \frac{c^2_n}{k^2+m^2_n},~~
 G^{\{(M),(MM)\}}= \sum^\infty_{n=0}\frac{\{c_nc_n^{(M)},
 (c_n^{(M)})^2\}}{k^2+m^2_n}\label{19a}\ee

 Finally  the total $G_{ab} (k)= \delta_{ab} G(k)$ describing  the $q\bar q$ PS system with chiral field
 to the second order is written as
 \be
 G(k) =-\frac{N_c}{2} \frac{\Psi(k)}{(k^2+ m^2_\pi) \Phi(k)}, ~~
 \Phi(k)
 =\sum^\infty_{n=0}\frac{(c_n^{(M)})^2}{(k^2+m^2_n)(m^2_n-m^2_\pi)}.
 \label{20}\ee

 One can see in (\ref{20}) the pion pole given by GOR relation
 (\ref{19}) and separated from the quark model dynamics, while the
 $q\bar q$ (quark model) poles at  $k^2=-m^2_n$ are shifted into a
 new position, e.g. the first pion radial excitation in shifted
 down
 \be k^2=-m^2_1(1+\delta_1) , m^2_1 \delta_1 =
 -\frac{c^2_1(m_1^2-m_0^2)(m_0^2-m^2_\pi)}{c_1^2(m_0^2-m^2_\pi)+c_0^2(m^2_1-m^2_\pi)}.
 \label{21}\ee
 Resultingly using string Hamiltonian in [19] it was found that
 shifts are
 $$ \pi(1S),~~ m_0=0.51~{\rm GeV}~\to m'_0=0.14~{\rm
 GeV~~(exact)}$$
$$ \pi(2S),~~ m_1=1.51~{\rm GeV}~\to m'_1=1.25~{\rm
 GeV~} (\exp: 1.3~{\rm GeV})~$$
$$ \pi(3S),~~ m_2=2.18~{\rm GeV}~\to m'_2=1.98~{\rm
 GeV~} (\exp: 1.8~{\rm GeV})~$$
and similar results for kaon radial excitations \cite{19}.

\section{Using QML for decay and channel coupling}

The QML, Eq. (\ref{10}) with the chiral degrees of freedom to  the
lowest order can be  rewritten as
 \be
 \Delta L^{(1)}= \int \overline{\psi}(x)\sigma |\vex|\gamma_{5}
 \frac{\pi^{A}\lambda^{A}}{F_{\pi}} \psi(x) dt d^{3}x \label{22} \ee
 Using Dirac equation for $\psi(x), \bar \psi(x)$ one can connect
 (\ref{22}) with the standard Weinberg  Lagrangian \cite{2}, also
 expanded to the lowest order
  \be
 \Delta L^{ch} = g^{q}_{A} tr (\overline\psi\gamma_{\mu} \gamma_{5}
 \omega_{\mu} \psi),\omega =\frac{i}{2F_{\pi}}(u \partial_\mu u^{+}-
 u^{+} \partial_{\mu} u),~~ u=\sqrt{\hat U} \label{23} \ee
 where the constant $g^q_A$ should be  taken as $g^q_A=1$. This Lagrangian was tested for pionic transitions successfully
in \cite{21}. On the other hand  the Lagrangian (\ref{22}) was
suggested in \cite{22} as the  basic Lagrangian  for chiral decays
of mesons and baryons. This decay Lagrangian couples only quark
and chiral degrees of freedom  and physically corresponds to the
case when pion  (kaon)  is emitted or absorbed by the quark at the
end of the string. Note that QML (\ref{22}) does not contain any
fitting  parameters and   therefore produces  unambigious
predictions for decay amplitudes. Recently \cite{23} it was used
to calculate the shift of the mass  of $D_s(J^P=0^+)$   due to the
coupling to  the  channel DK. The resulting shift is expressed in
terms of overlap  integral of the  $s$-quark wave-functions and
constant $f_k\cong 1.2 f_\pi=1.2 \cdot 94$ MeV, and was found to
be equal to 100 MeV, in good agreement with experimental mass of
2317 MeV.

\section{Conclusions}

It is shown that the same bilinear  field correlator $\lan F(x)
F(y)\ran \sim D(x-y)$,\\ which is  responsible for confinement
produces also CSB with the chiral fields entering  simply as a
factor $\exp (i \hat \phi \gamma_5)$ multiplying the  scalar
confining potential $\sigma |\ver|$. Therefore CSB and confinement
should disappear at the same temperature  $T_c$, in agreement with
lattice data \cite{24}. This combined chiral-confining dynamics
naturally explains pattern of meson spectra, numerical values of
$f_\pi$ and  quark condensate and decay  transitions.\\

{\bf \Large  Acknowledgements}\\

Support from  the Federal Program of the Russian Ministry of
Industry, Science and Technology No.40.052.1.1.1112 and  from the
grant for scientific schools NS-1774. 2003. 2 is gratefully
acknowledged.\\

\end{document}